\documentclass[12pt]{article}
\usepackage[margin=0.7in]{geometry}
\usepackage{float}
\usepackage{amsmath}
\usepackage{amssymb}
\usepackage{graphicx}
\usepackage{cite}
\usepackage{color}
\usepackage{dcolumn}
\usepackage{bm}
\usepackage{booktabs}
\usepackage{newfloat}
\DeclareFloatingEnvironment[name={SI Figure}]{suppfigure}
\DeclareFloatingEnvironment[name={SI equation}]{suppequation}
\DeclareFloatingEnvironment[name={SI Table}]{supptable}
\usepackage{hyperref}
\begin{document}

\begin{flushleft}
{\LARGE
\textbf{Multilayer network decoding versatility and trust}
}
\\
Camellia Sarkar$^{1}$, Alok Yadav$^{2}$ \& Sarika Jalan$^{1,2,\ast}$
\\ 
\it ${^1}$ Centre for Biosciences and Biomedical Engineering, Indian Institute of Technology Indore, Simrol, Indore 452020, India\\
\it ${^2}$ Complex Systems Lab, Discipline of Physics, Indian Institute of Technology Indore, Simrol, Indore 452020, India\\

$\ast$ Corresponding author e-mail: sarika@iiti.ac.in
\end{flushleft}

\begin{abstract}
In the recent years, the multilayer networks have increasingly been realized as a more realistic framework to understand emergent physical phenomena in complex real world systems. We analyze a massive
time-varying social data drawn from the largest film industry of the world under multilayer network framework. The framework enables us to evaluate the versatility of actors, which
turns out to be an intrinsic property of lead actors. Versatility in dimers suggests that working with different types of nodes are more beneficial than with similar ones. However, the triangles yield a different relation between type of co-actor and the success of lead nodes indicating the importance of higher order motifs in understanding the properties of the underlying system. Furthermore, despite the degree-degree correlations of entire networks being neutral, multilayering picks up different values of correlation indicating positive connotations like trust, in the recent years. Analysis of weak ties of the industry uncovers nodes from lower degree regime being important in linking Bollywood clusters. The framework and the tools used herein may be used for unraveling the complexity of other real world systems.
\end{abstract}

\section{Introduction}
Complex network science revolves around the hypothesis that the behavior of complex systems can be explained in terms of the structural and functional relationships between their components by means of a graph representation \cite{Barabasi_Rev_2002}. The network framework provides cue into whether the structural environment confers opportunities for or constraints on individual action \cite{Borgatti_2009}. Social network formation is a complex process in which individuals create and deactivate social ties in order to simultaneously satisfy their goals under multiple (possibly conflicting) constraints. Dynamics of social behaviors ranging from opinions, cultural and linguistic traits, crowd
behavior, hierarchy formation, social spreading and human dynamics and their connections have been investigated using various tools of statistical physics \cite{Santo}. Essential to understanding the behavior of humans within their socio-economical environment is the
observation that they simultaneously play different roles in various interconnected social networks, such
as friendship networks, communication networks, family, or business networks \cite{Szell_Soc_Net,Cozzo}. 
 This superposition of a number of networks is often termed as multilayer networks \cite{multilayer1,Domenico_2}.
In a multilayer network, on one hand, actions of individuals help in defining the topological structure of networks, and on the other hand, the topology constrains and
shapes the possible actions of individuals \cite{multilayer2,Battison,Bianconi} leading to a spurt in the activities of modeling real world complex systems and behavior, for example,
multimodal transportation networks \cite{multilayer_transport}, climatic systems \cite{multilayer_climate}, the human brain \cite{multilayer_brain} and failure and robustness \cite{multilayer_energy}.
It has recently been realized that neglecting the multilayer structure leads to wrong identification of the most versatile nodes, overestimating the importance of more marginal agents \cite{Domenico_NatComm}.

In this study, we use a huge empirical social data drawn from the largest film industry in the world, Bollywood.
This film industry has gained worldwide coverage by selling an estimated 3.6 billion tickets as compared to Hollywood's 2.6 billion as per CBFC 2006 statistics \cite{Lorenzen}.
Furthermore, it is said to reflect or affect the decisions and preferences of the populace \cite{Bose_book}. 
On one hand, Bollywood is driven by the underlying practices prevalent in the society, while on the other hand strongly influences the mass, thus acting as a mirror of the society  \cite{FICCI}. Moreover, network studies on Bollywood data have demonstrated how changes in interactions of the underlying networks are associated with events occurring in the society \cite{SJ_social}. 
Based on all the above, this film industry can be referred to as a model to understand and predict various aspects of society based on complex interaction patterns among the members.

We construct the Bollywood networks as follows: the nodes are the actors and interactions between a pair of actors is defined if they have co-acted in a movie. We assort the co-actor data comprising of movies and their corresponding actors. Owing to the rapidly changing nature of the society \cite{Mishra_Bollywood}, we divide the data into intervals of five years.
This time frame on one hand is small enough to capture the minute changes shaping in the society and on the other hand is large enough so as to not miss out any prolific change occurring in the society. We take three such datasets in order to see how the properties of the networks change with time. Considering genres as layers, we construct multilayer networks for each time span. Our analysis using multilayer network approach, on one hand, reveals the importance of versatility of the nodes for their success, while on the other hand indicates the significance of interactions between all types of nodes in the model system. Emergence of cooperation with time is also revealed through our investigations. 

\section{Methods}
{\bf Curation of data and network construction:} We collect the Bollywood data from a reputed movie repository website \cite{repository}. We extract the names of all the movies, their sequential star cast list
and genre information from 1998 to 2012 and segregate them into three datasets, each consisting of data for five years. 
We curate the movies and their details under 33 different genres. Many of the genres
had very few movies and considering them separately would have yielded statistically insignificant results. Thus, we combine the genres having very few
movies and categorize them as `Others'. This gives us eight broad genres namely `Action', `Comedy', `Drama', `Romance', `Thriller', `Crime and Horror', `Social' and `Others'. We treat each genre as a layer of the multilayer social network. For a particular layer, a pair of actors $i$ and $j$ are connected if they have co-acted in a movie of the corresponding genre in the respective dataset. Thus, we obtain eight different sub-networks for each multilayer network corresponding the three datasets. For each layer $\alpha$, the elements of the adjacency matrix, $A^\alpha$ is given as
\begin{equation}
A_{\mathrm {ij}}^\alpha = \begin{cases} 1~~\mbox{if } i \sim j \\
0 ~~ \mbox{otherwise} \end{cases},
\end{equation}
All the adjacency matrices are symmetric (i.e. $A_{ij}^\alpha$ = $A_{ji}^\alpha$). The degree of an actor $d_i^\alpha$ is the number of actors that actor has worked with, given as $d_i^\alpha$ = $\sum_{j=1}^{N} A_{ij}^\alpha$. Note that not every actor has worked in movies of every genre. Thus the number of nodes may differ across the genres. The lead male actors can be defined as the actors occupying the first position in the movie star cast providing an easy way of segregating lead actors and supporting actors. However, it may be possible that supporting actors have acted as lead actors in few movies appearing in the first position of the star cast \cite{repository}. Further, it is also possible that some actors have acted in only one or two movies, but in lead roles occupying the first position in star cast. In order to avoid such cases, we require a threshold for
finding lead actors, which is a common
practice used for construction of networks from real data, for instance in construction of gene
coexpression networks \cite{gene_threshold}. Note that a very low value of threshold may bring in lot of the supporting actors in the list of lead actors, whereas a very high value of threshold may leave behind many known lead actors \cite{SM}. After conducting several trials
with different threshold values, we find five movies to be the optimal criteria for all the three datasets.
This broadly divides the actors into two types, one consisting of the lead male actors (denoted by L) and the other comprising of the rest of the actors, which includes both supporting actors and female actors (denoted by S).

{\bf Filmfare data assimilation:} Of the different awards categories introduced over the years of cinematic heritage in Bollywood, the Filmfare awards, voted by the public and a committee of experts is one of the oldest and the most reputed one \cite{Filmfare_reputed}. We extract the Filmfare award nominations data from the website chronologically \cite{Filmfare} and count the number of times
every actor is nominated in each five-year span. Instead of the awards bagged we rather take into
account the award nominations in order to avoid the interplay of some kind of bias affecting the decision.

{\bf Dimers and triangles in one or more layers:} If a dimer, a pair of interacting nodes, is present in only one of the eight layers, it is defined as a dimer unique to one layer. We such dimers are called dimers unique to one layer. If a dimer is present in only any two of the layers, we call it a dimer unique to two layers. Similarly, dimers unique to higher layers are defined. Further, if a triangle, a complete subgraph of order three, is present in only one of the eight layers, it is defined as a triangle unique to one layer. If a triangle is present in only any two of the layers, we call it a triangle unique to two layers. In a similar manner, triangles unique to higher layers are defined. Note that every lead actor appearing in unique LL and LS dimers or LLL, LLS, LSS and SSS triangles is counted only once, even if the lead actor has appeared in more than one unique LL, LS dimer or or LLL, LLS, LSS and SSS triangle.

{\bf Structural measures:}
We quantify the degree-degree correlations of a network by considering the Pearson (degree-degree) correlation coefficient, given as \cite{Newman_assortativity}
 \begin{equation}
\label{assortativity}
r = \frac{[N_{c}^{-1}\sum_{l=1}^{N_{c}} {d_{i}}^{l} {d_{j}}^{l}] - [N_{c}^{-1}\sum_{l=1}^{N_{c}} \frac{1}{2}({d_{i}}^{l} + {d_{j}}^{l})^{2}]}
{[N_{c}^{-1}\sum_{l=1}^{N_{c}} \frac{1}{2}({d_{i}^{l}}^{2}+ {d_{j}^{l}}^{2})] - [N_{c}^{-1}\sum_{l=1}^{N_{c}}\frac{1}{2}({d_{i}}^{l} + {d_{j}}^{l})^{2}]}, 
 \end{equation}
where $d_i^l$, $d_j^l$ are the degrees of nodes at both the ends of the $l^{th}$ connection and $N_c$ represents the total connections in the network.

In order to investigate the hypothesis of weak ties on networks \cite{Onnela}, we calculate the link betweenness centrality and overlap as follows. Link betweenness centrality for an undirected link $e$ is defined as $\beta_{L} = \sum_{v \in V_{s}} \sum_{w \in V/{v}} \sigma_{vw} (e)/\sigma_{vw}$,
where $\sigma_{vw} (e)$ is the number of shortest paths between $v$ and $w$ that contain $e$, and $\sigma_{vw}$ is the total number of shortest paths between $v$
and $w$.
The overlap of the neighborhood of two connected
nodes $i$ and $j$ is defined as $O_{ij} = \frac{n_{ij}}{(d_{i} - 1) + (d_{j} - 1) - n_{ij}}$,
where $n_{ij}$ is the number of neighbors common to both nodes $i$ and $j$. Here $d_i$ and $d_j$ represent the degree of the $i^{th}$ and $j^{th}$ nodes.

\section{Results}
{\bf Structural properties of individual layers:} 
The degree distribution of many of the layers follow power law, evaluated using maximum likelihood estimation \cite{Clauset} (values of parameters provided in Table S1 of \cite{SM}).
Rest of them resemble power law behavior, but the maximum likelihood method rejects the power law hypothesis.
Further, we find that there is drastic increase in the size of the networks with time. In fact from 98-02 to 08-12, the size of the network becomes almost double (Table~\ref{all}).
However, despite a drastic increase in size, the average connectivity of the nodes (actors) remain almost same across the three time spans (Table~\ref{all}) indicating the average connectivity as an intrinsic property of the model system.

Further, the average clustering coefficient ($\langle C \rangle$), which measures the average connectivity of the neighbors \cite{Watts}, depict much higher values for all the three datasets (Table~\ref{Properties_multilayer}) as compared those of the corresponding random networks (Table S2 of \cite{SM}), which is quite expected as various complex social and biological networks are known to exhibit high average clustering coefficient \cite{Barabasi_Rev_2002}.
\begin{table}
\caption{The properties of all the three Bollywood datasets.
Here $N$, $N_c$, $\langle k \rangle$ and $r$ represent the size, number of connections, average degree and overall assortativity coefficient value of the network, respectively.}
\label{all}
\begin{center}
\begin{tabular}{|c|c|c|c|c|}    \hline
\footnotesize{Dataset}     & \footnotesize{$N$} & \footnotesize{$N_c$} & $\langle k \rangle$ & \footnotesize{$r$} \\ \hline
\footnotesize{08-12} 	& \footnotesize{3934} &	\footnotesize{59698} & \footnotesize{30} & \footnotesize{0.04} 	\\ \hline 
\footnotesize{03-07} 	& \footnotesize{3041}	& \footnotesize{55345} & \footnotesize{36} & \footnotesize{0.07}	 \\ \hline 
\footnotesize{98-02}	& \footnotesize{1899}	& \footnotesize{44277} & \footnotesize{47} & \footnotesize{-0.03}	  \\ \hline 
\end{tabular}
\end{center}
\end{table}

{\bf (Dis)likelihood in connectivity uncovers cooperation in the recent times:} 
The (dis)assortativity has emerged as an important structural measure, used for understanding (dis)likelihood in connectivity in the underlying systems \cite{Newman_assortativity}. Various
social networks are known to be assortative, while many of
the biological and technological networks are found to be
disassortative \cite{Assortativity_ref}.
We calculate the values of assortativity coefficient using Eq. \ref{assortativity} for the three networks without consideration of genres and find that all the networks exhibit a value of assortativity coefficient close to zero (Table \ref{all}). Based on the rapidly changing nature of the society \cite{Mishra_Bollywood}, we may expect some changes reflected in the interaction behavior of the model system. But correlation in degrees without consideration of genres does not exhibit any change across time.
However, in the following, we show how multilayer approach demonstrates changes in the interaction patterns with time.
The networks corresponding to individual layers exhibit different values of assortativity coefficients
(Table \ref{Properties_multilayer}). The same actors working in different genres may contribute differently to the assortativity coefficients of respective genres based on the degrees of their co-actors in each genre. Further, we generate an ensemble of corresponding random networks of the same size and same degree sequence \cite{configuration} as of the Bollywood layers. The $r$ values of different layers in the three time spans (Table \ref{Properties_multilayer}) are significantly larger than their values in corresponding random networks (Table S3 of \cite{SM}).

We find that the number of genres exhibiting assortative nature increase from the older to the latest dataset (Table \ref{Properties_multilayer}), though the entire networks without consideration of genres show neutral nature of degree correlations as discussed above. This is interesting in the light that networks with links having positive connotations like trust and endorsement have been shown to exhibit assortative mixing in degrees, while disassortativity has been related with negative connotations like disapproval and distrust \cite{Signed_social}. The increase in the number of genres exhibiting assortativity, as we proceed from 98-02 to 08-12 dataset, can be related to the enhancement of positive connotations among the actors with time. This along with the fact that Bollywood has become more successful in the recent years \cite{Lorenzen,Bollywood_success} might be an indication that trust is beneficial for success. This change in the network structure is also revealed in the next section while we are investigating the frequency of occurrence of actors in different genres under multilayer framework, relating intrinsic properties with their individual success.
\begin{figure}
\centerline{\includegraphics[width=0.6\linewidth]{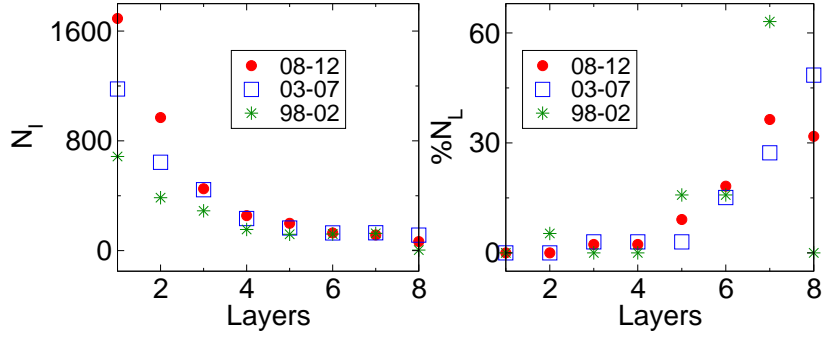}}
\caption{Actors uniquely common in one or more layers in 08-12, 03-07 and 98-02 datasets. $N_{I}$ refers to the number of actors, uniquely common in one or more layers. $\%N_{L}$ represents the percentage of lead actors out of the $N_{I}$ actors, who are uniquely common in one or more layers.}
\label{Fig2}
\end{figure}

{\bf Interplay of versatility and success:} 
The way we construct the multilayer networks allows us to capture one more important property, i.e. versatility. Versatility, in the present context, can be referred to as the number of genres a particular actor has worked in \cite{Domenico_NatComm,Latora}. It is known to be a qualitative feature of actors \cite{actor_versatility} and entrepreneurs \cite{entrepreneur_versatility} and has been emphasized as a property of the ones excelling in multiple dimensions \cite{versatility}. 
We find that the number of actors which includes all the lead and the supporting actors unique to one or more layers decrease with an increase in the number of layers (Fig.~\ref{Fig2}). What stands interesting is that the number of lead actors, accounting to $\sim1.2\%$ of the total actors, that are unique to one or more number of layers consistently increase with an increase in the number of layers, which means that 
versatility is an intrinsic property of lead actors. It is noteworthy that none of the lead actors are found to be confined to only one layer (Fig.~\ref{Fig2}) indicating that versatility is essential for lead actors. Such importance of versatility has also been reported in the biological systems, where the essential genes constituting a very small fraction of the total human genome encode proteins which have been known to participate in various signaling pathways \cite{Essential_genes}, which can be considered different layers. Akin to lead actors from the successful Bollywood industry \cite{Bollywood_success}, these essential genes are very few in number \cite{essential_genes_conserved}. The importance of these essential genes in various signaling pathways can be related to the significance of versatility in the success of lead actors. 
\begin{figure}
\centerline{\includegraphics[width=0.6\linewidth]{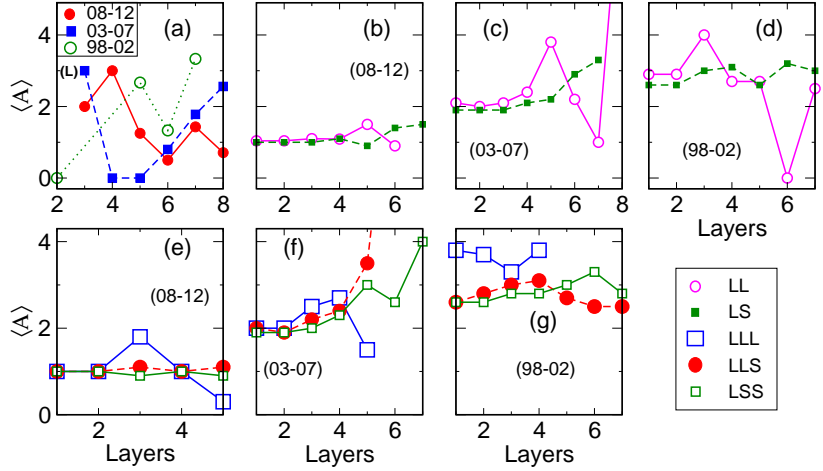}}
\caption{The average number of award nominations plotted as a function of number of layers (a) for lead actors (represented as L) unique to one or more layers in 08-12, 03-07 and 98-02 datasets; (b), (c) and (d) for lead actors participating in unique LL (lead-lead) and LS (lead-supporting) dimers in 08-12, 03-07 and 98-02 datasets, respectively; (e), (f) and (g) for lead actors participating in unique LLL (lead-lead-lead), LLS (lead-lead-supporting) and LSS (lead-supporting-supporting) triangles in 08-12, 03-07 and 98-02 datasets, respectively.}
\label{Fig3}
\end{figure}
\begin{table*}
\caption{The properties of different layers of the Bollywood multilayer networks, 08-12 represented by `i' superscript, 03-07 represented by `ii' superscript and 98-02 represented by `iii' superscript having size 3934, 3041 and 1899 nodes, respectively.
Here $N_c$, $N$, $\langle k \rangle$, $\langle C \rangle$ and $r$, respectively, represent the number of connections, number of participating nodes, average degree,
average clustering coefficient and assortativity coefficient for each layer. Note that the social genre has $r$ value 1 in the 98-02 dataset owing to only five movies in that genre. None of the actors have acted in more than one movie in that genre. This gives rise to five complete subgraphs rendering $r$ value one.}
\label{Properties_multilayer}
\begin{center}
\begin{tabular}{|c|c|c|c|c|c|c|c|c|c|c|c|c|c|c|c|}    \hline
\footnotesize{Layer}&\footnotesize{$N^{i}$}&\footnotesize{${N_c^{i}}$}&\footnotesize{${\langle k\rangle^{i}}$}&\footnotesize{${\langle C \rangle^{i}}$}&
\footnotesize{$r^{i}$}&\footnotesize{$N^{ii}$}&\footnotesize{${N_c^{ii}}$}&\footnotesize{${\langle k\rangle^{ii}}$}&\footnotesize{${\langle C \rangle^{ii}}$}&
\footnotesize{$r^{ii}$}&\footnotesize{$N^{iii}$}&\footnotesize{${N_c^{iii}}$}&\footnotesize{${\langle k\rangle^{iii}}$}&\footnotesize{${\langle C \rangle^{iii}}$}&
\footnotesize{$r^{iii}$}\\\hline
\footnotesize{Social}&\footnotesize{1543}&\footnotesize{18048}&\footnotesize{23}&\footnotesize{0.8}&\footnotesize{-0.1}&\footnotesize{887}&\footnotesize{9891}&\footnotesize{22}&\footnotesize{0.9}&\footnotesize{0.1}&\footnotesize{65}&\footnotesize{632}&\footnotesize{19}&\footnotesize{1}&\footnotesize{1}\\\hline
\footnotesize{Drama}&\footnotesize{1375}&\footnotesize{14003}&\footnotesize{20}&\footnotesize{0.9}&\footnotesize{0.1}&\footnotesize{1456}&\footnotesize{21905}&\footnotesize{30}&\footnotesize{0.8}&\footnotesize{-0.1}&\footnotesize{1094}&\footnotesize{20713}&\footnotesize{38}&\footnotesize{0.8}&\footnotesize{0}\\\hline
\footnotesize{Comedy}&\footnotesize{1345}&\footnotesize{17232}&\footnotesize{26}&\footnotesize{0.8}&\footnotesize{0}&\footnotesize{1013}&\footnotesize{13865}&\footnotesize{27}&\footnotesize{0.8}&\footnotesize{-0.1}&\footnotesize{550}&\footnotesize{8160}&\footnotesize{30}&\footnotesize{0.8}&\footnotesize{0}\\\hline
\footnotesize{Romance}&\footnotesize{1121}&\footnotesize{9821}&\footnotesize{18}&\footnotesize{0.9}&\footnotesize{-0.1}&\footnotesize{1085}&\footnotesize{13819}&\footnotesize{26}&\footnotesize{0.8}&\footnotesize{-0.1}&\footnotesize{808}&\footnotesize{15050}&\footnotesize{37}&\footnotesize{0.8}&\footnotesize{0}\\\hline
\footnotesize{Thriller}&\footnotesize{946}&\footnotesize{7296}&\footnotesize{15}&\footnotesize{0.9}&\footnotesize{0.1}&\footnotesize{766}&\footnotesize{7658}&\footnotesize{20}&\footnotesize{0.8}&\footnotesize{-0.1}&\footnotesize{392}&\footnotesize{4685}&\footnotesize{24}&\footnotesize{0.8}&\footnotesize{0}\\\hline
\footnotesize{Action}&\footnotesize{688}&\footnotesize{6796}&\footnotesize{20}&\footnotesize{0.9}&\footnotesize{0.1}&\footnotesize{672}&\footnotesize{9435}&\footnotesize{28}&\footnotesize{0.8}&\footnotesize{-0.1}&\footnotesize{666}&\footnotesize{12018}&\footnotesize{36}&\footnotesize{0.8}&\footnotesize{0}\\\hline
\footnotesize{Crime}&\footnotesize{556}&\footnotesize{3985}&\footnotesize{14}&\footnotesize{0.9}&\footnotesize{0.2}&\footnotesize{511}&\footnotesize{6948}&\footnotesize{27}&\footnotesize{0.9}&\footnotesize{0.2}&\footnotesize{455}&\footnotesize{6149}&\footnotesize{27}&\footnotesize{0.8}&\footnotesize{0}\\\hline
\footnotesize{Others}&\footnotesize{1850}&\footnotesize{13400}&\footnotesize{15}&\footnotesize{0.8}&\footnotesize{0.1}&\footnotesize{1766}&\footnotesize{21139}&\footnotesize{24}&\footnotesize{0.8}&\footnotesize{0.2}&\footnotesize{1185}&\footnotesize{19940}&\footnotesize{34}&\footnotesize{0.7}&\footnotesize{0}\\\hline
\end{tabular}
\end{center}
\end{table*}
  
While the versatility of lead actors does not show any definite relation with their success, captured by the number of award nominations (Fig.~\ref{Fig3}(a)), interestingly investigation of versatility in motifs exhibit relation with the properties of individual nodes forming these motifs.
Since the nodes are of two types, namely L and S, various network motifs can be drawn on the basis of different combinations of these nodes, such as LL and LS dimers, which are motifs of order two and LLL, LLS, LSS and SSS triangles, which are motifs of order three. Similar to the versatility of individual nodes, considering multilayer network approach, we define the versatility of motifs as the number of layers they appear. In the following, we show how existence of two types of nodes in the versatile motifs provides us a relation between the success of lead actors and the type of actors they work with. We find that the average number of award nominations of lead actors remains largely independent of the type of node they are paired up with (Figs.\ref{Fig3}(b), (c) and (d)). This behavior is found to be the same across the three datasets. However, when actor dimers are present in more than six layers, the lead actors in LS dimers emerge more successful as compared to those in LL dimers (Figs.~\ref{Fig3}(b), (c) and (d)). This indicates that interactions between dissimilar types of nodes is more beneficial when these interacting partners are versatile. Only one LL dimer in 03-07 dataset defies this trend owing to the exceptional success of the actor pair in the film industry (discussed in supplementary material \cite{SM}). 

Further, we investigate the relation between versatility in triangles and the success of lead nodes. We find that the success behavior of lead actors in triangles is different from that of those in dimers. As plotted in Figs.~\ref{Fig3}(e), (f) and (g), when triangles are present in less number of layers, the average number of award nominations of lead actors in unique LLL triangles are higher than those in unique LLS and LSS triangles. Although, as discussed, appearing in LL dimers does not considerably affect the success of lead actors, appearing in LLL triangles stands beneficial for their success. Furthermore, the relation between versatility of triangles and success of lead actors for LLS and LSS triangles is different from that followed by LLL triangles (Figs.~\ref{Fig3}(e), (f) and (g)). This indicates that the higher order interaction patterns can have different behavior as compared to lower order.
Note that the standard deviations of award nominations are quite high for many of the datasets (Table S4 of \cite{SM}), indicating the existence of very high as well as very low award nominations in those datasets. These large standard deviations indicate that on average, not all the lead actors in LS category are more successful than all the lead actors in LL category, but at least one or few of them together in LS category have more award nominations than one or few of them together in LL category. Just to have a unified measure for assessing the success, we consider the average award nominations.

The analysis done here so far reveals the relation of the success behavior of individuals with the changes in the interaction patterns between two different types of the nodes.
In the following, analysis of the properties of links reveals a third category of nodes which help in linking different Bollywood clusters.
\begin{figure}
\centerline{\includegraphics[width=0.6\linewidth]{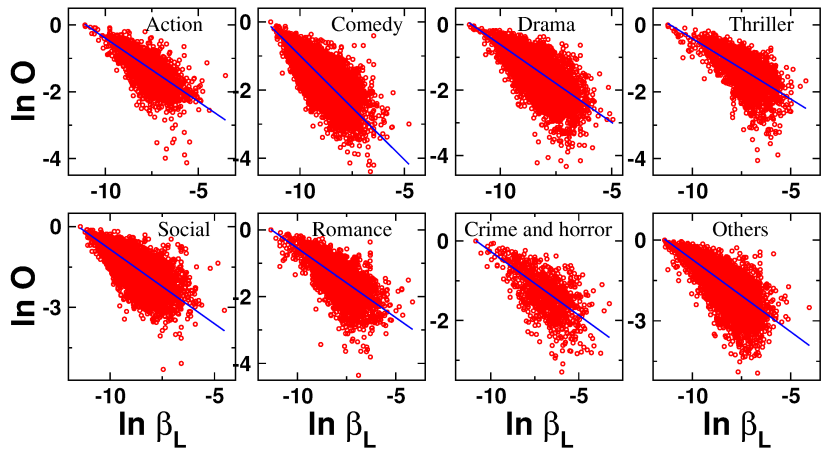}}
\caption{Overlap as a function of link betweenness centrality on a logarithmic scale fitted with a straight line for the 08-12 dataset.}
\label{Fig4}
\end{figure}

{\bf Importance of every node:} 
An important proposition of sociology is the `Weak ties hypothesis' which was initially proposed by Granovetter \cite{Granovetter}. We use this concept in the context of networks, where the ties having low overlap in their neighborhoods (i.e. less number of common neighbors) are termed as the weak ties as per Onnela \cite{Onnela}.
These weak ties have high link betweenness centrality and are the ones known to bridge different communities \cite{Onnela,Szell_Soc_Net}. In our study, we find that the overlap of ties exhibits negative correlation with their link betweenness centrality (Fig.~\ref{Fig4}), i.e. the ties with less overlap have tendency to have high link betweenness centrality.
We then investigate the properties of the end nodes of the ties having low overlap and high link betweenness centrality and find that these nodes are actors who are mostly neither lead actors, defined based on their position of appearance in the movie star cast, nor they are popular supporting actors (high degree nodes). In fact, these nodes appear in lower degree regime and can be considered to form a third category of the nodes. Details of correlation values of overlap and link betweenness centrality, and list of actors forming weak ties with high link betweenness centrality along with their degrees are provided in supplementary material \cite{SM}. Earlier works \cite{SJ_social} having already emphasized the importance of high and moderate degree nodes, this result here picks up relatively lower degree nodes which are important in linking different Bollywood clusters, reflecting the importance of all types of the nodes.

\section{Conclusion}
Various real world complex systems encompass multiple types of interactions \cite{Multiplex_new} and hence the multilayer networks provide a better framework for investigating the underlying properties of such systems. We find an increase in the number of assortative genres in the film industry with time indicating increase in trust among interacting partners \cite{Signed_social}. Since this data is drawn from a system which has gained progressive success over the years \cite{Bollywood_success}, positive connotations can be considered a factor underlying its success. Using the multilayer framework, we evaluate the frequency of occurrence of actors in different layers, termed as versatility, which emerges as an intrinsic property of lead actors.
Furthermore, apart from the existence of various genres representing layers, the presence of two types of nodes provides a richer framework for investigating the versatility of motifs.
Our analysis reveals that the success behavior of lead actors appearing in unique triangles is different from those working in dimers, which suggests that higher order motifs can provide a different insight about a system. The results here are drawn based on dimers and triangles as analysis of higher order motifs comprising of lead actors would not yield significantly comparable results since the number of lead actors are very less. Motifs are known to be the basic building blocks of various social, technological and biological networks \cite{Alon_motif}, hence this mode of analysis can be extended to other systems which have considerably high number of (dis)similar nodes, for instance predator-prey networks \cite{predator-prey1} in order to identify the factors governing the stability of these natural communities \cite{predator-prey2}. Furthermore, the analysis of the weak ties reveal the involvement of lower degree nodes in holding the system together, thus demonstrating the importance of all types of nodes which are defined here based on their degrees. 

While in this study, we segregate the movie co-actor data into layers based on genres, the segregation of genres based on the type of emotion, either positive or negative, can further yield interesting results. Such investigations on the relation between types of neighboring nodes and individual properties in genres with positive and negative emotions can be one of the aspects of future investigation and might attract research in disciplines like psychology \cite{Psychology}.

\section{Acknowledgements}
SJ acknowledges CSIR grant (25(0205)/12/EMR-II) and DST grant (EMR/2014/000368) for the financial 
support. CS thanks Sanjiv Kumar Dwivedi, Amit Kumar Pawar and Animesh Chaturvedi for their help in assimilating the movie database and writing various codes. AY acknowledges CSIR for his fellowship grant.

\newpage

\cleardoublepage
{\Large \bf Supplementary Material}
\vspace{1cm}
\begin{supptable}[h]
\begin{center}
\caption{Significance of power law fitting for all the subnetworks in 08-12, 03-07 and 98-02 datasets.}
\begin{tabular}{|l|l|l|l|l|l|l|l|l|l|l|l|l|l|l|l|}
\toprule 
    Layers & \multicolumn{5}{|c|}{08-12} & \multicolumn{5}{|c|}{03-07} & \multicolumn{5}{|c|}{98-02} \\ \hline
    & \footnotesize{$\gamma$} & \footnotesize{$x_{min}$} & \footnotesize{p-value} & \footnotesize{KS} & \footnotesize{$A/R$}
    & \footnotesize{$\gamma$} & \footnotesize{$x_{min}$} & \footnotesize{p-value} & \footnotesize{KS} & \footnotesize{$A/R$}
    & \footnotesize{$\gamma$} & \footnotesize{$x_{min}$} & \footnotesize{p-value} & \footnotesize{KS} & \footnotesize{$A/R$}  \\
    \midrule
 \footnotesize{Social}	&	\footnotesize{2.88} & \footnotesize{25} & \footnotesize{0} & \footnotesize{0.05} & \footnotesize{R}	&	\footnotesize{3.5}	&	\footnotesize{37}	&	\footnotesize{0.04}	&	\footnotesize{0.07}	& \footnotesize{close} & \footnotesize{1.68}	&	\footnotesize{4}	&	\footnotesize{0}	&	\footnotesize{0.28}	&	\footnotesize{R}\\ \hline
\footnotesize{Drama}	&	\footnotesize{3.5} & \footnotesize{44} & \footnotesize{0.36} & \footnotesize{0.05} & \footnotesize{A}&	\footnotesize{3.09}&	\footnotesize{37}	&	\footnotesize{0}	&	\footnotesize{0.06} & \footnotesize{close} & \footnotesize{2.86}	&	\footnotesize{38}	&	\footnotesize{0}	&	\footnotesize{0.06}	&	\footnotesize{R}\\ \hline
\footnotesize{Comedy}	&	\footnotesize{2.67}	&	\footnotesize{30}	&	\footnotesize{0.05}	&	\footnotesize{0.04}	&	\footnotesize{A}	&	\footnotesize{3.2}	&	\footnotesize{45}	&	\footnotesize{0.12}	& \footnotesize{0.06} & \footnotesize{A} & \footnotesize{3.5} & \footnotesize{59} & \footnotesize{0.79} & \footnotesize{0.06} & \footnotesize{A}	\\ \hline
\footnotesize{Romance}	&	\footnotesize{2.81}	&	\footnotesize{13}	&	\footnotesize{0}	&	\footnotesize{0.06}	&	\footnotesize{R}	&	\footnotesize{2.65}	&	\footnotesize{20}	&	\footnotesize{0}	& \footnotesize{0.05}	& \footnotesize{R} & \footnotesize{2.98} & \footnotesize{48} & \footnotesize{0} & \footnotesize{0.08} & \footnotesize{R}	\\ \hline
\footnotesize{Thriller}	&	\footnotesize{3.5}	& 	\footnotesize{30}	&	\footnotesize{0.05}	&	\footnotesize{0.08}	&	\footnotesize{A}	&	\footnotesize{3.5}	&	\footnotesize{29}	&	\footnotesize{0.1}	&	\footnotesize{0.06} & \footnotesize{A} & \footnotesize{3.5} & \footnotesize{40} & \footnotesize{0.53} & \footnotesize{0.06} & \footnotesize{A}	\\ \hline
\footnotesize{Action}	&	\footnotesize{3.5}	&	\footnotesize{37}	&	\footnotesize{0.3}	&	\footnotesize{0.06}	&	\footnotesize{A}	&	\footnotesize{3.5}	&	\footnotesize{50}	&	\footnotesize{0.07}	&	\footnotesize{0.08} & \footnotesize{A} & \footnotesize{3.23} & \footnotesize{53} & \footnotesize{0} & \footnotesize{0.08} & \footnotesize{R}	\\ \hline
\footnotesize{Crime}&\footnotesize{2.93}&\footnotesize{11}&\footnotesize{0}&\footnotesize{0.06}&\footnotesize{R}&\footnotesize{3.5}&\footnotesize{49}&\footnotesize{0.1}&\footnotesize{0.13}&\footnotesize{A}&\footnotesize{3.5}&\footnotesize{40}&\footnotesize{0.03}&\footnotesize{0.08}&\footnotesize{close}\\\hline
\footnotesize{Others}&\footnotesize{3.5}&\footnotesize{52}&\footnotesize{0.1}&\footnotesize{0.08}&\footnotesize{A}&\footnotesize{3.5}&\footnotesize{65}&\footnotesize{0.02}&\footnotesize{0.07}&\footnotesize{close}&\footnotesize{3.5}&\footnotesize{100}&\footnotesize{0.13}&\footnotesize{0.07}&\footnotesize{A}\\
    \bottomrule
\end{tabular}
\end{center}
\label{powerlaw}
\end{supptable}

We evaluate the significance of power law in the different Bollywood subnetworks of three time spans by using maximum-likelihood fitting methods with goodness-of-fit tests based on the Kolmogorov-Smirnov (KS) statistic and likelihood ratios. The power-law hypothesis is accepted if the p-value is $\geq$ 0.05.

\begin{supptable}[h]
\begin{center}
\caption{Average clustering coefficient values for corresponding ER random networks}
\begin{tabular}{|l|l|l|l|}	\hline
Layer 	& $\langle C \rangle_{08-12}^{ER}$	& $\langle C \rangle_{03-07}^{ER}$   & $\langle C \rangle_{98-02}^{ER}$  \\ \hline
Social	&	0.02	&	0.02	&	0.3	\\ \hline
Drama	&	0.01	&	0.02	&	0.04	\\ \hline
Comedy	&	0.02	&	0.03	&	0.06	\\ \hline
Romance	&	0.02	&	0.02	&	0.05	\\ \hline
Thriller	&	0.02	&	0.03	&	0.06	\\ \hline
Action	&	0.03	&	0.04	&	0.05	\\ \hline
Crime	&	0.03	&	0.05	&	0.06	\\ \hline
Others	&	0.01	&	0.01	&	0.03	\\ \hline
\end{tabular}
\end{center}
\label{random_C}
\end{supptable}

\begin{supptable}[h]
\begin{center}
\caption{Pearson (degree-degree) correlation coefficient values for corresponding random networks having the same degree sequence as of the Bollywood networks, known as configuration model.}
\begin{tabular}{|l|l|l|l|}	\hline
Layer 	&  $r_{08-12}^{conf}$	& $r_{03-07}^{conf}$   & $r_{98-02}^{conf}$  \\ \hline
Social	&	0.02		&	-0.01	&	-0.05\\ \hline
Drama	&	-0.03	&	0.01		&	0.02\\ \hline
Comedy	&	0.02		&	-0.02	&	0.02\\ \hline
Romance	&	-0.01	&	-0.03	&	0.01\\ \hline
Thriller	&	-0.03	&	-0.03	&	0.04\\ \hline
Action	&	-0.02	&	0		&	-0.02\\ \hline
Crime	&	0.02		&	0.02		&	-0.03\\ \hline
Others	&	-0.02	&	0.01		&	0\\ \hline
\end{tabular}
\end{center}
\label{random_r}
\end{supptable}

{\bf \underline{Threshold for enlisting lead actors}}: A very low value of threshold brings in lot of the supporting actors in the list of lead actors, whereas a very high value of threshold may leave behind many known lead actors. For 08-12 dataset, decreasing the threshold from 5 to 4 movies increases the count of lead actors from 44 to 62, while increasing the threshold to 6 movies decreases the count to 27. For 03-07 dataset, decreasing the threshold from 5 to 4 movies increases the count of lead actors from 33 to 45, while increasing the threshold to 6 movies decreases the count to 24. For 98-02 dataset, decreasing the threshold from 5 to 4 movies increases the count of lead actors from 19 to 27, while increasing the threshold to 6 movies decreases the count to 15.

\begin{supptable}[h]
\begin{center}
\caption{Standard deviation of award nominations for dimers and triangles in 08-12, 03-07 and 98-02 datasets.}
\begin{tabular}{|l|l|l|l|l|l|l|l|l|l|l|l|l|l|l|l|}
\toprule 
    $N_{Layers}$ & \multicolumn{5}{|c|}{08-12} & \multicolumn{5}{|c|}{03-07} & \multicolumn{5}{|c|}{98-02} \\ \hline
    & \footnotesize{LL} & \footnotesize{LS} & \footnotesize{LLL} & \footnotesize{LLS} & \footnotesize{LSS}
    & \footnotesize{LL} & \footnotesize{LS} & \footnotesize{LLL} & \footnotesize{LLS} & \footnotesize{LSS}
    & \footnotesize{LL} & \footnotesize{LS} & \footnotesize{LLL} & \footnotesize{LLS} & \footnotesize{LSS}  \\
    \midrule
\footnotesize{1}	&	\footnotesize{1.2} 	& 	\footnotesize{1.1} 	& 	\footnotesize{2.9} 	& 	\footnotesize{2.8} 	& 	\footnotesize{2.8}	&	\footnotesize{2.9}	&	\footnotesize{2.8}	&	\footnotesize{1.2}	&	\footnotesize{1.1} 	& \footnotesize{1.1} 	& \footnotesize{2.6}	& \footnotesize{2.4}	& \footnotesize{2.5}	&\footnotesize{2.4}	& \footnotesize{2.4}\\ \hline
\footnotesize{2}	&	\footnotesize{1.1}	&	\footnotesize{1.1}	&	\footnotesize{2.9}	&	\footnotesize{2.8}	&	\footnotesize{2.8}	&	\footnotesize{2.9}	&	\footnotesize{2.8}	&	\footnotesize{1.2}	& 	\footnotesize{1.1} 	& \footnotesize{1.1} 	& \footnotesize{2.5} 	& \footnotesize{2.4} 	& \footnotesize{2.6} 	& \footnotesize{2.4} 	& \footnotesize{2.4}	\\ \hline
\footnotesize{3}	&	\footnotesize{1.2}	&	\footnotesize{1.2}	&	\footnotesize{3.1}	&	\footnotesize{2.9}	&	\footnotesize{2.8}	&	\footnotesize{2.9}	&	\footnotesize{2.8}	&	\footnotesize{1.4}	& 	\footnotesize{1.2}	& \footnotesize{1.1} 	& \footnotesize{2.2} 	& \footnotesize{2.4} 	& \footnotesize{1.5} 	& \footnotesize{2.5} 	& \footnotesize{2.4}	\\ \hline
\footnotesize{4}	&	\footnotesize{1.3}	& 	\footnotesize{1.2}	&	\footnotesize{3.3}	&	\footnotesize{3.1}	&	\footnotesize{3.1}	&	\footnotesize{3}	&	\footnotesize{2.9}	&	\footnotesize{0.9}	&	\footnotesize{0.9} 	& \footnotesize{1.2} 	& \footnotesize{2.5} 	& \footnotesize{2.4} 	& \footnotesize{3} 	& \footnotesize{2.6} 	& \footnotesize{2.4}	\\ \hline
\footnotesize{5}	&	\footnotesize{1.6}	&	\footnotesize{1.2}	&	\footnotesize{0}	&	\footnotesize{3.6}	&	\footnotesize{3.5}	&	\footnotesize{3.6}	&	\footnotesize{3.1}	&	\footnotesize{0.6}	&	\footnotesize{0.8} 	& \footnotesize{0.7} 	& \footnotesize{2.6} 	& \footnotesize{2.5} 	& \footnotesize{-} 	& \footnotesize{2.6} 	& \footnotesize{2.4}	\\ \hline
\footnotesize{6}	&	\footnotesize{0.8}	&	\footnotesize{1.4}	&	\footnotesize{-}	&	\footnotesize{2.8}	&	\footnotesize{3.9}	&	\footnotesize{2.7}	&	\footnotesize{3.4}	&	\footnotesize{-}	&	\footnotesize{-}	&\footnotesize{-}	&\footnotesize{-}	&\footnotesize{2.6}	&\footnotesize{-}	&\footnotesize{0.7}	&\footnotesize{2.6}\\\hline
\footnotesize{7}	&	\footnotesize{-}	&	\footnotesize{0.7}	&	\footnotesize{-}	&	\footnotesize{-}	&	\footnotesize{0}	&	\footnotesize{0}	&	\footnotesize{3.7}	&	\footnotesize{-}	&	\footnotesize{-}	&\footnotesize{-}	&\footnotesize{0.7}	&\footnotesize{2.1}	&\footnotesize{-}	&\footnotesize{0.7}	&\footnotesize{2.6}\\
\footnotesize{8}	&	\footnotesize{-}	&	\footnotesize{-}	&	\footnotesize{-}	&	\footnotesize{-}	&	\footnotesize{-}	&	\footnotesize{2.8}	&	\footnotesize{-}	&	\footnotesize{-}	&	\footnotesize{-} 	& \footnotesize{-} 	& \footnotesize{-} 	& \footnotesize{-} 	& \footnotesize{-} 	& \footnotesize{-} 	& \footnotesize{-}	\\ \hline    
\end{tabular}
\end{center}
\label{powerlaw}
\end{supptable}

\begin{supptable}[!ht]
\begin{center}
\caption{High degree nodes in 08-12 dataset. The table shows the top two degree nodes naming hub 
actors in each layer followed by their ranking in multiplex network ($R_{new}$) 
and the networks studied before ($R_{old}$).}
\begin{tabular}{|c|c|c|c|c|c|c|c|}         \hline 
Layer       & Hub Actors & $R_{new}$  & $R_{old}$  \\ \hline
Action           & Mushtaq Khan, Yashpal Sharma  & 1,2	& 4, 8	  \\ 
Crime $\&$ horror & Zakir Hussain, Murli Sharma   & 1,2	& 11, 10  \\ 
Comedy           & Rajpal Yadav, Manoj Joshi     & 1,2	& 6, 12	  \\ 
Drama            & Anupam Kher, Boman Irani      & 1,2	& 1, 27	  \\ 
Other            & Govind Namdev, Jackie Shroff  & 1,2	& 7, 13	  \\ 
Romance          & Anupam Kher, Boman Irani      & 1,2 	& 1, 27	  \\ 
Social           & Rajpal Yadav, Anupam Kher     & 1,2	& 6, 1	  \\ 
Thriller         & Zakir Hussain, Gulshan Grover & 1,2 	& 11, 20  \\ \hline 
\end{tabular}
\label{Table 2} 
\end{center}
\end{supptable}
\begin{supptable}[!ht]
\begin{center}
\caption{High degree nodes in 03-07 dataset. The table shows the top two degree nodes naming hub 
actors in each layer followed by their ranking in multiplex network ($R_{new}$) 
and the networks studied before ($R_{old}$).}
\begin{tabular}{|c|c|c|c|c|c|c|c|}         \hline 
Layer       & Hub Actors & $R_{new}$  & $R_{old}$  \\ \hline
Action           & Gulshan Grover, Suniel Shetty, Rajpal Yadav  & 1,2,2	& 4, 5, 3	  \\ 
Crime $\&$ horror & Pratima Kazmi, Suresh Dubey   & 1,2	& 34, 199  \\ 
Comedy           & Rajpal Yadav, Asrani     & 1,2	& 3, 26	  \\ 
Drama            & Amitabh Bachchan, Om Puri      & 1,2	& 1, 12	  \\ 
Other            & Rani Mukherjee, Amitabh Bachchan  & 1,2	& 25, 1	  \\ 
Romance          & Satish Shah, Sharat Saxena      & 1,2 	& 21, 7	  \\ 
Social           & Anupam Kher, Vivek Shauq     & 1,2	& 2, 19	  \\ 
Thriller         & Gulshan Grover, Priyanka Chopra & 1,2 	& 4, 37  \\ \hline 
\end{tabular}
\label{Table 5} 
\end{center}
\end{supptable}
\begin{supptable}[!ht]
\begin{center}
\caption{High degree nodes in 98-02 dataset. The table shows the top two degree nodes naming hub 
actors in each layer followed by their ranking in multiplex network ($R_{new}$) 
and the networks studied before ($R_{old}$).}
\begin{tabular}{|c|c|c|c|c|c|c|c|}         \hline 
Layer       & Hub Actors & $R_{new}$  & $R_{old}$  \\ \hline
Action           & Johny Lever, Mohan Joshi  & 1,2	& 1, 8	  \\ 
Crime $\&$ horror & Johny Lever, Razzak Khan   & 1,2	& 1, 3  \\ 
Comedy           & Johny Lever, Razzak Khan     & 1,2	& 1, 3	  \\ 
Drama            & Johny Lever, Kulbhushan Kharbanda      & 1,2	& 1, 16	  \\ 
Other            & Shakti Kapoor, Anil Nagrath  & 1,2	& 2, 5	  \\ 
Romance          & Johny Lever, Alok Nath      & 1,2 	& 1, 10	  \\ 
Social           &                             & 1,2	& 	  \\ 
Thriller         & Johny Lever, Dinesh Hingoo & 1,2 	& 1, 4  \\ \hline 
\end{tabular}
\label{Table 8} 
\end{center}
\end{supptable}

\begin{supptable}[t]
\begin{center}
\caption{Pair-wise analysis of actors uniquely common to one, two or more layers in 08-12, 03-07 and 98-02 datasets. $N_{pair}$ denotes the number of pairs that are uniquely common to one or more layers and $r_{pair}$ gives the average of the degree-degree correlation of these actor pairs. 
$P_{LL}$, $P_{LS}$ and $P_{SS}$ represent the number of lead-lead, lead-supporting and supporting-supporting actor pairs uniquely common to one, two or more layers.}
\begin{tabular}{|c|c|c|c|c|c|c|c|c|c|c|}    \hline
Layers & $N_{pair}^{08-12}$ & $P_{LL}^{08-12}$ &$N_{pair}^{03-07}$ & $P_{LL}^{03-07}$&$N_{pair}^{98-02}$ & $P_{LL}^{98-02}$\\\hline
1 & 33696 &111 &25692  &101&42616  &13 \\\hline
2 & 20542 & 96&16415 & 54&10680 & 17     \\ \hline
3 &3718 & 43&8707 &36&7056 & 5    \\ \hline
4 &  901 & 19&2514 & 35&2986 & 9	   \\ \hline
5 & 158 & 8&1668 &13&1322 & 7	   \\  \hline
6 &  34 & 4&180 & 7&391 &- 	   \\ \hline
7 &  7 & -&	69 & 1&119 & 1   \\ \hline
8 &  - & - & 7 &1& - &- 	   \\ \hline
\end{tabular}
\begin{flushleft}
\end{flushleft}
\label{Common}
\end{center}
\end{supptable}
\begin{supptable}[t]
\begin{center}
\caption{Number of cliques ($\Delta$) unique to one, two or more layers for 08-12 dataset and $\Delta_I$ denotes participating nodes. $\Delta_{LLL}$, $\Delta_{LLS}$, $\Delta_{LSS}$ and  $\Delta_{SSS}$ denote number of cliques unique to LLL, LLS, LSS and SSS cliques, respectively.}
\begin{tabular}{|c|c|c|c|c|c|c|c|c|c|c|}    \hline
Layers &$N_{\Delta}$ & $\Delta_I$ & $\Delta_{LLL}$ & $\Delta_{LLS}$ & $\Delta_{LSS}$ &  $\Delta_{SSS}$ \\ \hline
1 & 2,03,253 &2983 & 206 & 5187 & 43412 & 154448  \\ \hline 
2 & 1,23,177 &1719 & 86 & 2438 & 25851 & 94802  \\ \hline 
3 & 13,525 &566&4 & 200 & 2545 & 10776  \\ \hline 
4 & 2,118 &171 & 2 & 52 & 401 & 1663  \\ \hline 
5 & 67 &276 &1 & 15 & 23 & 28  \\ \hline 
6 & - &- & - & - & - & -  \\ \hline 
7 & 1 &3 &- & - & - & 1  \\ \hline 
8 & - & - & - & - & - & -  \\ \hline 
\end{tabular}
\begin{flushleft}
\end{flushleft}
\label{Common_cliques_08-12}
\end{center}
\end{supptable}
\begin{supptable}[b]
\begin{center}
\caption{Number of cliques unique in one, two or more layers for 03-07 dataset. The terminologies are the same as used in 
Table~\ref{Common_cliques_08-12}.}
\begin{tabular}{|c|c|c|c|c|c|c|}    \hline
Layers &$N_{\Delta}$ & $\Delta_I$ & $\Delta_{LLL}$ & $\Delta_{LLS}$ & $\Delta_{LSS}$ &  $\Delta_{SSS}$ \\ \hline
1 & 2,10,962 & 2216 &786 & 9497 & 48548 & 152131  \\ \hline 
2 & 1,24,310 &1338 & 111 & 2219 & 19454 & 102526  \\ \hline 
3 & 57,531 &818 & 34 & 742 & 7690 & 49065  \\ \hline 
4 & 12,710 &361 & 19 & 412 & 2973 & 9306  \\ \hline 
5 & 12,035 & 202 &1 & 63 & 1841 & 10130  \\ \hline 
6 & 136 & 50 &- & 4 & 57 & 75  \\ \hline 
7 & 42 & 17&- & - & 3 & 39  \\ \hline 
8 & 3 & 1 &- & - & - & 1  \\ \hline 
\end{tabular}
\begin{flushleft}
\end{flushleft}
\label{Common_cliques_03-07}
\end{center}
\end{supptable}
\begin{supptable}[t]
\begin{center}
\caption{Number of cliques unique in one, two or more layers for 98-02 dataset. The terminologies are the same as used in 
Table~\ref{Common_cliques_08-12}.}
\begin{tabular}{|c|c|c|c|c|c|c|}    \hline
Layers &$N_{\Delta}$ & $\Delta_I$ & $\Delta_{LLL}$ & $\Delta_{LLS}$ & $\Delta_{LSS}$ &  $\Delta_{SSS}$ \\ \hline
1 & 2,27,222& 1470 & 22 & 1579 & 33159 & 192462  \\ \hline 
2 & 1,07,170& 933 & 6 & 714 & 16184 & 90266  \\ \hline 
3 & 70,134& 627 & 1 & 281 & 9740 & 60112  \\ \hline 
4 & 17,659& 344 & 2 & 130 & 3341 & 14186  \\ \hline 
5 & 4,836& 223 & - & 48 & 1008 & 3780  \\ \hline 
6 & 358 & 109& - & 1 & 66 & 291  \\ \hline 
7 & 83 & 35 & - & 1 & 10 & 72  \\ \hline 
8 & - & - & - & - & - & -  \\ \hline 
\end{tabular}
\begin{flushleft}
\end{flushleft}
\label{Common_cliques_98-02}
\end{center}
\end{supptable}

{\bf \underline{Exception in LL dimers}}: Note that in the 03-07 dataset, there is only one LL actor pair Amitabh Bachchan - Abhishek Bachachan the father-son duo who have appeared in all the eight genres. It has already been discussed in our earlier work that Amitabh Bachchan has always had a successful realm and visibly stands out of the domain of our analysis and defies the common trends. Taking this aspect into consideration, we do not consider this LL actor pair in our analysis. 

\begin{suppfigure}
\centering
\includegraphics[width=0.49\columnwidth]{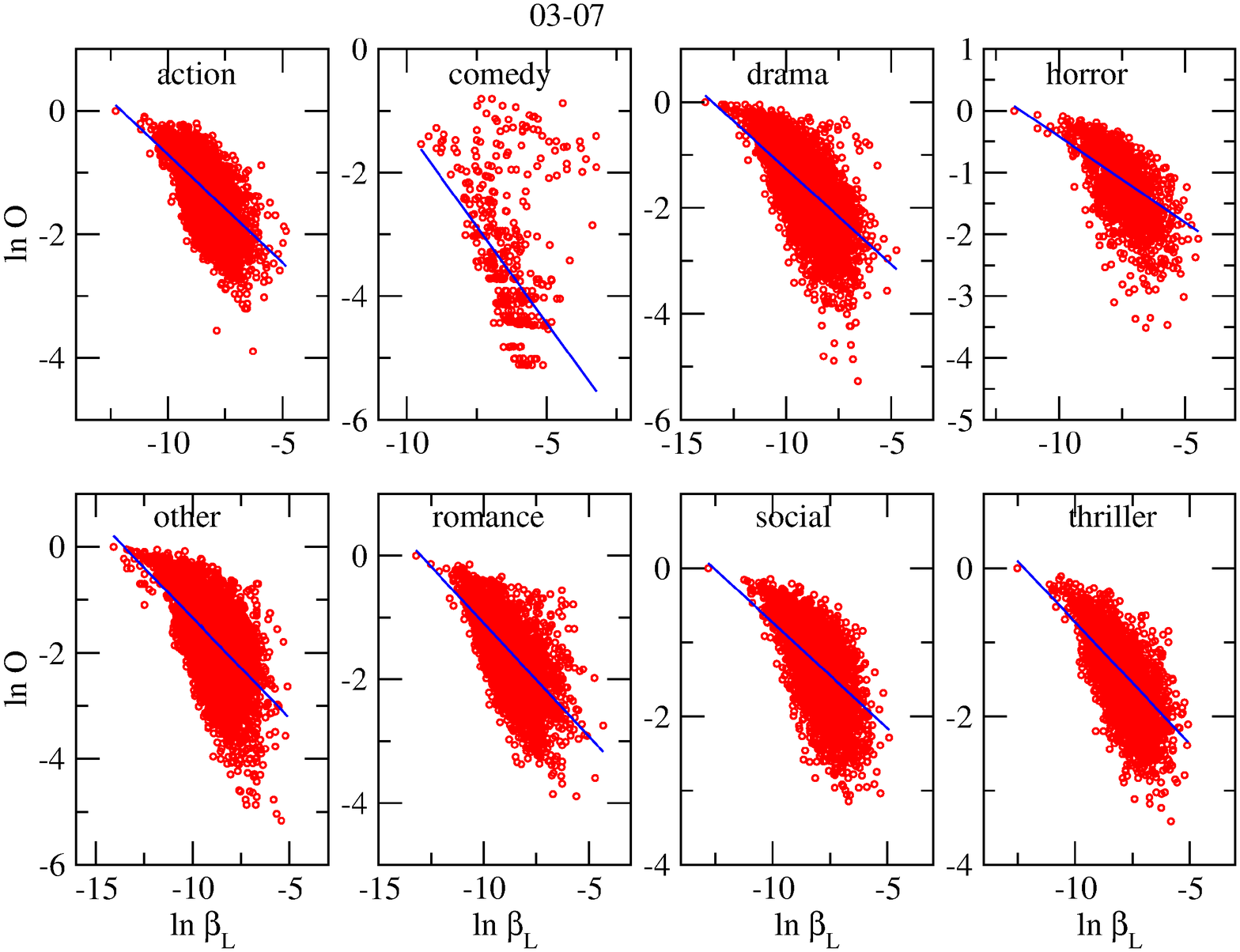}
\includegraphics[width=0.49\columnwidth]{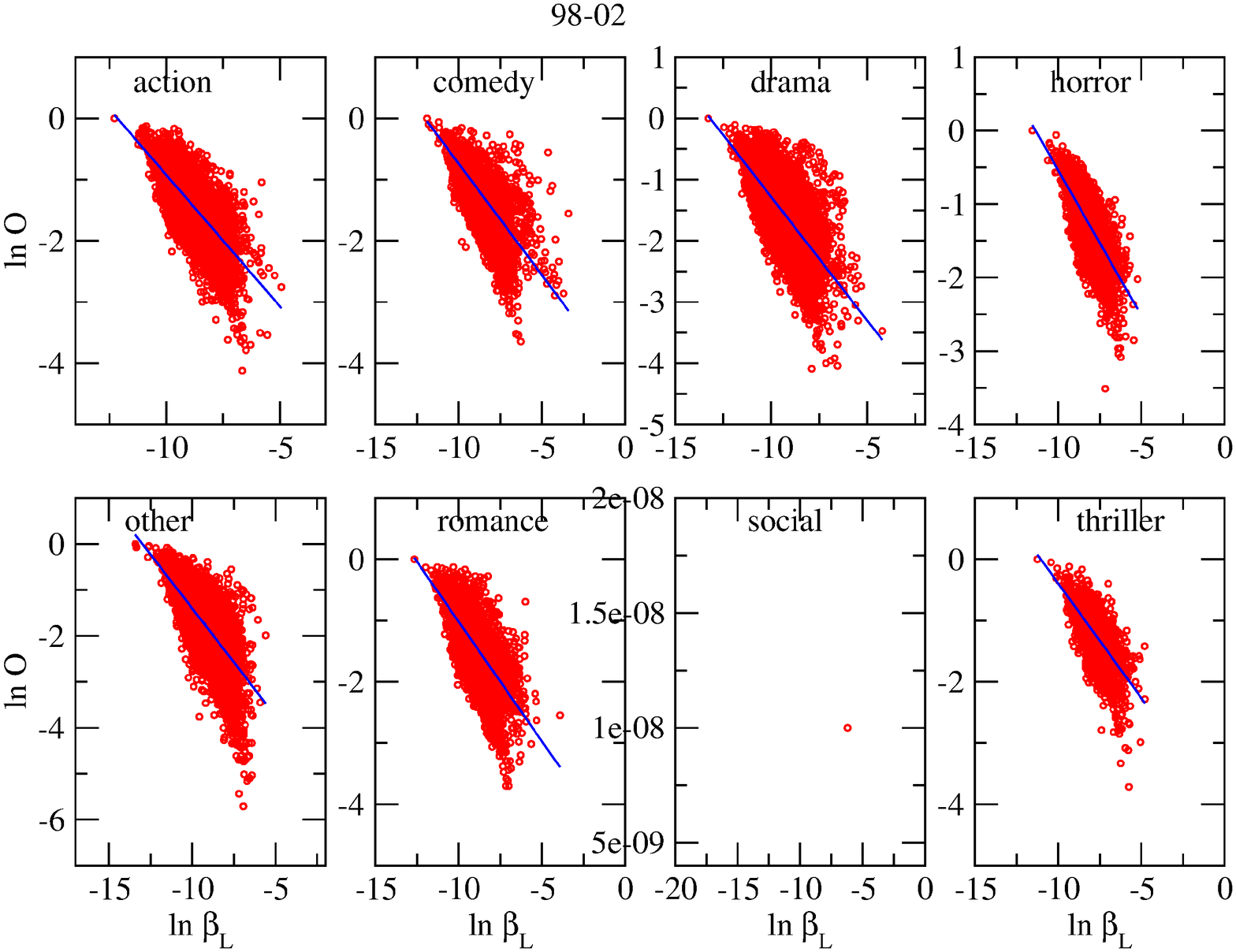}
\caption{Overlap versus link betweenness centrality on a logarithmic scale for 03-07 and 98-02 datasets.}
\label{overlap_link_bet}
\end{suppfigure}

\begin{supptable}[t]
\begin{center}
\caption{List of actor-pairs having high link betweenness centrality but low overlap for 08-12 dataset.}
\begin{tabular}{|c|c|c|}    \hline
{\bf Layer} & {\bf $\beta_{L}$-$O_{corr}$} &{\bf Links (High $\beta_{L}$ low $O$)} \\ \hline
Action & -0.48 &Amit Behl (107) - Dalip Tahil (246) \\ \hline 
Comedy & -0.43 &Aslam Khan (55) - Ankita Shrivastava (10)  \\ \hline 
Drama & -0.58 & Om Puri (418) - Emil Marwa (16)  \\ \hline 
Thriller & -0.46 &Rahul Bose (111) - Nandita Das (43)  \\ \hline 
Social & -0.50 &Naseeruddin Shah (254) - Iman Ali (15)  \\ \hline 
Romance &-0.49 & Anushka Sharma (46) - Parineeta Chopra (24) \\ \hline 
Crime and Horror &  -0.44 &Hemant Pandey (195) - Paintal (72)  \\ \hline 
Others & -0.57 & Anil Nagrath (134) - Vinod Tripathi (23)\\ \hline 
\end{tabular}
\begin{flushleft}
\end{flushleft}
\label{Overlap_link_bet_08-12}
\end{center}
\end{supptable}

\begin{supptable}[t]
\begin{center}
\caption{List of actor-pairs having high link betweenness centrality but low overlap for 03-07 dataset.}
\begin{tabular}{|c|c|c|}    \hline
{\bf Layer} & {\bf $\beta_{L}$-$O_{corr}$} & {\bf Links (High $\beta_{L}$ low $O$)} \\ \hline
Action & -0.52 & Gulshan Grover (389) - Govinda (158) \\ \hline 
Comedy & -0.009 & Delnaz Paul (74) - Rajpal Yadav (434)  \\ \hline 
Drama & -0.41 & Esha Deol (196) - Meera Jasmine (22)  \\ \hline 
Thriller & -0.55 & Gulshan Grover (389) - Manisha Koirala (116)  \\ \hline 
Social & -0.59&  Yashpal Sharma (256) - Kashish Duggal (22)  \\ \hline 
Romance & -0.36 & Dinesh Hingoo (264) - Shweta Menon (158) \\ \hline 
Crime and Horror & -0.64 & Zakir Hussain (186) - Dilip Prabhawalkar (119)  \\ \hline 
Others & -0.41 & Cheran (7) - Prakash Raj (57)  \\ \hline 
\end{tabular}
\begin{flushleft}
\end{flushleft}
\label{Overlap_link_bet_03-07}
\end{center}
\end{supptable}

\begin{supptable}[t]
\begin{center}
\caption{List of actor-pairs having high link betweenness centrality but low overlap for 98-02 dataset.}
\begin{tabular}{|c|c|c|}    \hline
{\bf Layer} & {\bf $\beta_{L}$-$O_{corr}$} &{\bf Links (High $\beta_{L}$ low $O$)} \\ \hline
Action & -0.44 & Mukesh Rishi (254) - Soundarya (18)  \\ \hline 
Comedy &  -0.25&  Simran (97) - Nagesh (23)  \\ \hline 
Drama & -0.34 &Tabu (229) - Sukumari (43)  \\ \hline 
Thriller &  -0.60 & Johny Lever (571) - Sanjay Mishra (63) \\ \hline 
Social & - & -  \\ \hline 
Romance & -0.37 & Ameesha Patel (114) - Pawan Kalyan (13) \\ \hline 
Crime and Horror & -0.60 & Anil Nagrath (405) - Paresh Rawal (331)  \\ \hline 
Others & -0.39 & Tabu (229) - Gajraj Rao (64) \\ \hline 
\end{tabular}
\begin{flushleft}
\end{flushleft}
\label{Overlap_link_bet_98-02}
\end{center}
\end{supptable}

\end{document}